\documentclass[twocolumn,prb,superscriptaddress,showpacs,amsmath,amssymb]{revtex4}

\usepackage{graphicx}

\newcommand{\UNA}{\ensuremath{\rm UNi_2Al_3}}
\newcommand{\UPA}{\ensuremath{\rm UPd_2Al_3}}
\newcommand{\YRS}{\ensuremath{\rm YbRh_2Si_2}}
\newcommand{\Tc}{\ensuremath{T_\textrm{c}}}

\begin{document}

\title{Microwave spectroscopy on heavy-fermion systems: probing the dynamics of charges and magnetic moments}

\author{Marc Scheffler}
\email[]{scheffl@pi1.physik.uni-stuttgart.de}
\affiliation{1.\ Physikalisches Institut, Universit\"at Stuttgart, D-70550 Stuttgart, Germany}

\author{Katrin Schlegel}
\affiliation{1.\ Physikalisches Institut, Universit\"at Stuttgart, D-70550 Stuttgart, Germany}

\author{Conrad Clauss}
\affiliation{1.\ Physikalisches Institut, Universit\"at Stuttgart, D-70550 Stuttgart, Germany}

\author{Daniel Hafner}
\affiliation{1.\ Physikalisches Institut, Universit\"at Stuttgart, D-70550 Stuttgart, Germany}

\author{Christian Fella}
\affiliation{1.\ Physikalisches Institut, Universit\"at Stuttgart, D-70550 Stuttgart, Germany}

\author{Martin Dressel}
\affiliation{1.\ Physikalisches Institut, Universit\"at Stuttgart, D-70550 Stuttgart, Germany}

\author{Martin Jourdan}
\affiliation{Institut f\"ur Physik, Johannes Gutenberg Universit\"at, D-55099 Mainz, Germany}

\author{J\"org Sichelschmidt}
\affiliation{Max-Planck-Institut f\"ur Chemische Physik fester Stoffe, D-01187 Dresden, Germany}

\author{Cornelius Krellner}
\affiliation{Max-Planck-Institut f\"ur Chemische Physik fester Stoffe, D-01187 Dresden, Germany}
\affiliation{Physikalisches Institut, Goethe-Universit\"at Frankfurt, D-60438 Frankfurt/Main, Germany}

\author{Christoph Geibel}
\affiliation{Max-Planck-Institut f\"ur Chemische Physik fester Stoffe, D-01187 Dresden, Germany}

\author{Frank Steglich}
\affiliation{Max-Planck-Institut f\"ur Chemische Physik fester Stoffe, D-01187 Dresden, Germany}

\date{\today}

\begin{abstract}

Investigating solids with light gives direct access to charge dynamics, electronic and magnetic excitations. For heavy fermions, one has to adjust the frequency of the probing light to the small characteristic energy scales, leading to spectroscopy with microwaves. We review general concepts of the frequency-dependent conductivity of heavy fermions, including the slow Drude relaxation and the transition to a superconducting state, which we also demonstrate with experimental data taken on \UPA. We discuss the optical response of a Fermi liquid and how it might be observed in heavy fermions.
Microwave studies with focus on quantum criticality in heavy fermions concern the charge response, but also the magnetic moments can be addressed via electron spin resonance (ESR). We discuss the case of \YRS, the open questions concerning ESR of heavy fermions, and how these might be addressed in the future.
This includes an overview of the presently available experimental techniques for microwave studies on heavy fermions, with a focus on broadband studies using the Corbino approach and on planar superconducting resonators.

\end{abstract}



\maketitle

\section{Introduction}\label{SectionIntroduction}

Heavy-fermion materials are prime examples of metals with electronic properties that are governed by strong electron-electron correlations. Though the simpler examples of heavy-fermion metals are well understood within theoretical frameworks based on the Kondo lattice, the wide variety of unusual properties observed in different heavy-fermion materials continuously draws the attention of experimental and theoretical physicists. Two particular interesting features are unconventional superconductivity and quantum criticality.\cite{Monthoux2007,Pfleiderer2009,vLoehneysen2007,Gegenwart2008} Here, heavy fermions have become model systems for other correlated electron system such as the cuprate superconductors. Concerning fundamental electronic properties, main advantages of heavy-fermion materials compared to other correlated metals when it comes to studies of their fundamental electronic properties are as follows: firstly, the characteristic energy scales of heavy fermions are comparably low, and thus they can be tuned by conveniently accessible values of pressure or magnetic field.
Secondly, the prime property of heavy fermions, namely their strongly enhanced effective mass, causes strong signatures in experimental observables such as the specific heat, the susceptibility, or the Fermi-liquid contribution to the transport scattering rate. This is even more the case because the low characteristic energy scales require that the experiments are performed at very low temperatures where other contributions to these observables that are not caused by the electronic system, but e.g.\ by phonons, are very weak. Therefore, the desired heavy-fermion response can be observed as a strong signal on top of a comparably weak background.
Thirdly, heavy-fermion materials often have a simple crystal structure, and in many cases single crystals of exceptional quality and very low residual resistivity can be grown.
These virtues of heavy fermions as model systems for correlated metals go hand in hand with an experimental challenge: the small characteristic energy scales require that experiments be performed at very low temperatures, on the scale of 1~K or even on the mK scale, i.e.\ in dilution refrigerators. Also, spectroscopic measurements have to be performed on these small energy scales, well below 1~meV (1~K $\approx$ 86~$\mu$eV $\approx$ 21~GHz).

These requirements can only be met by a few spectroscopic techniques. In particular photoemission spectroscopy, which otherwise is extremely helpful in understanding the electronic properties of correlated metals and which has been used successfully to study heavy-fermion materials at higher energies and temperatures,\cite{Klein2008} cannot directly access many of the most interesting heavy-fermion states because it does not have sufficient energy resolution and is incompatible with mK temperatures and the application of magnetic fields or pressure. 
Recently, scanning tunneling spectroscopy has managed to reach sufficiently low temperature and energy resolution. This has lead to studies of several heavy-fermion compounds at temperatures of a few K,\cite{Schmidt2010,Ernst2011,Aynajian2012} and corresponding studies at mK temperatures have become feasible.

Another rather direct energy-resolved access to the electronic properties of solids is optical spectroscopy.\cite{Dressel2002a,Basov2011} Incoming light directly interacts with the charges that are present in a sample, and therefore one can obtain information about the electronic properties of a material by measuring its optical properties. One big advantage of optical spectroscopy is that the probing energy can be adjusted to any relevant energy scale of a solid by choice of the appropriate spectral range. Furthermore, optical spectroscopy can be combined with low temperatures, high magnetic fields, and high pressure (although these parameters might complicate the experimental effort substantially). The most common spectral range to study correlated metals with optics is the infrared,\cite{Basov2011} typically covering photon energies 5~meV to 1~eV with Fourier transform spectrometers. For the next lower spectral range, THz techniques are used which typically have photon energies down to a few hundred $\mu$eV. To optically probe a material with photon energies of 100~$\mu$eV and below, microwave techniques have to be employed. Such experiments have to be performed in a fundamentally different manner than conventional optics: the wavelength of microwaves is so long (1~cm for frequency 30~GHz $\approx$ 124~$\mu$eV) that the light cannot be manipulated as beams in free space any more; furthermore a focal spot would be much larger than the typical sample size of heavy-fermion materials. Instead, the microwave signal is guided in coaxial cables or waveguides.

Like any electromagnetic radiation, microwaves can interact both with electric charges (via its electric field) and with magnetic moments (via its magnetic field). Optical spectroscopy of solids often neglects the magnetic component because the magnetic response usually is orders of magnitude weaker than the electrical one. The remainder of this paper will first discuss microwave spectroscopy on heavy fermions addressing the charge response and later microwave experiments that address the magnetic moments (via magnetic resonance).

\section{Microwave spectroscopy}

Microwave spec\-tros\-co\-py on metals at cryogenic temperatures has remained an experimental challenge. One reason is that for frequencies below the plasma edge, metals reflect almost 100\% of impinging light, and the difference to a reflectivity of unity is the quantity that has to be determined and contains the information about the material under study.\cite{Dressel2002a} With decreasing frequency, the reflectivity of a metal increases, and at microwave frequencies the reflectivity typically exceeds 99\%, which is very hard to measure with respect to a 100\% reference.
Another difficulty arises from the large wavelength of microwaves. This prevents the use of free-space propagation and windows to send the signal to a sample in a cryostat. Instead, the microwaves have to propagate in confining structures such as waveguides or coaxial cables. These cause substantial signal losses, which depend on frequency and temperature, and as such are difficult to calibrate during the experiment.

The traditional approach to overcome these difficulties are cavity resonators.\cite{Klein1993} If these have very high quality factors $Q$ even in the presence of the sample, then the enhanced interaction between microwave and sample (of order $Q$ compared to a single-bounce reflection measurement) leads to measurable effects even for samples with rather low losses. Furthermore, the relevant observables are the resonator frequency and the resonance linewidth, which are independent of the losses of the cables and therefore rather robust quantities.
Already since the 1980s, heavy-fermion materials have been studied with cavity resonators, and this has lead to the first experimental observations of the slow Drude response of heavy fermions at GHz frequencies.\cite{Beyermann1988a,Awasthi1993,Tran2002}
Cavity resonators are a very generic approach that allows microwave measurements on very different material classes, but there are two main drawbacks: firstly, because of geometric reasons (the sample is usually only a weak perturbation of the cavity fields, and the geometry of the sample in the cavity has to be known to great detail for data analysis) it is difficult to obtain precise absolute values of the microwave conductivity of the sample, in contrast to relative changes e.g.\ as a function of temperature or magnetic field.\cite{Tonegawa2012}
The second drawback is that with the traditional cavity approach one usually works at a single frequency only. Obtaining spectroscopic information requires a set of different cavities, and due to geometrical reasons, often also different samples.

These limitations have led to the development of new techniques. In the following, the two techniques that we employ for the study of heavy fermions will be discussed, namely broadband microwave spectroscopy in Corbino geometry and planar superconducting resonators. In addition, other microwave techniques have been developed in recent years that could be applied to heavy fermions as well, such as the bolometric technique (which is broadband and very sensitive, but does not give phase information)\cite{Turner2004} or multimode dielectric resonators (which can be compared to the planar superconducting resonators and which are particularly suited for experiments in high magnetic fields).\cite{Huttema2006}

\subsection{Broadband microwave spectroscopy}

Obtaining the full frequency dependence in a microwave experiment requires that all components, such as microwave source and detector, transmission lines as well as the interaction geometry of signal and sample, are truly broadband. Here the interaction of signal and sample can be achieved in two ways: firstly, the sample can form a part of a transmission line. For heavy-fermion materials this is difficult because they cannot be fabricated in the required extended geometries. The second possibility is a non-resonant single interaction such as transmission or reflection. 
Here, the so-called Corbino technique allows the study of the microwave conductivity of metals and superconductors at low temperatures.\cite{Booth1994} Its main idea is that the flat sample terminates a coaxial cable, i.e.\ it forms an electrical connection between inner and outer conductors. The microwave signal that travels along the coaxial cable is reflected by the sample, and the complex reflection coefficient, which can be measured by a network analyzer, directly contains the information about the sample properties. The difficulty of the Corbino technique at cryogenic temperatures is that the reflection coefficient is measured by the network analyzer at room temperature and thus also contains the damping and phase shift of the microwave signal that is not caused by the sample, but by the coaxial transmission line. These effects have to be calibrated with a rather arduous procedure.\cite{Booth1994,Stutzman2000a,Scheffler2005a}

The sensitivity of a Corbino experiment is not sufficient to resolve the properties of a highly conductive bulk sample such as a metal.\cite{Scheffler2005a} Therefore, one has to geometrically confine drastically the current-leading cross section of the sample. This can be done by reducing the sample thickness to much less than the skin depth of the material. In the case of metals, this means the use of thin films with thickness of order 100~nm. 
Unfortunately, it is rather difficult to grow high-quality thin films of heavy-fermion materials, and this is the reason why so far the only two heavy-fermion compounds studied in detail with Corbino spectroscopy are UPd$_2$Al$_3$ and UNi$_2$Al$_3$.\cite{Jourdan1999,Jourdan2004a} However, with recent success in the growth of Ce-based heavy-fermion thin films,\cite{Shishido2010,Mizukami2011,Li2011} more studies become foreseeable for the future.

Another development to further improve the sensitivity of Corbino measurements on conductive thin films is patterning the film in strip shapes.\cite{Scheffler2007,Steinberg2010} This can easily enhance the sensitivity by an additional factor of 5 to 10. Furthermore, this allows the study of anisotropic microwave properties for those materials where crystallographic directions within the film plane differ. In the case of heavy fermions, this can be obtained for UNi$_2$Al$_3$ films,\cite{Jourdan2004a,Zakharov2005} and here an anisotropic electrodynamic response was found already in THz transmission measurements that equally rely on thin-film samples.\cite{Ostertag2010,Ostertag2011}

\begin{figure}[t]%
\includegraphics*[width=\linewidth]{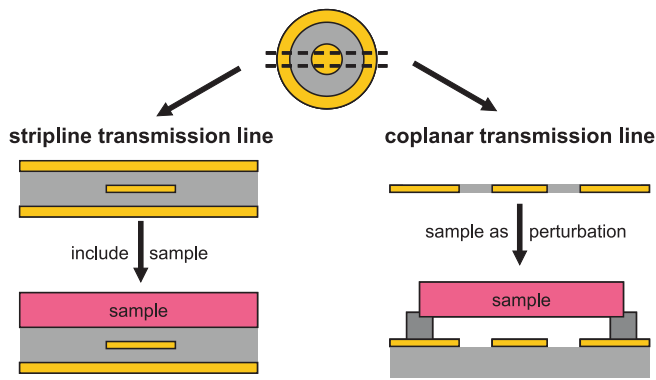}
\caption{%
  Schematic cross sections of microwave transmission lines. A stripline can be understood conceptionally as a coaxial cable squeezed flat, whereas a coplanar line can be understood as central cut through a coaxial cable. The metallic sample under study is one of the ground planes of the stripline or a perturbation of the cross-sectional geometry of the coplanar line.}
\label{Fig_CoaxToPlanarScheme}
\end{figure}

\subsection{Planar microwave resonators}\label{SectionPlanarResonators}

Because of limitations in sensitivity, the Corbino technique can resolve metallic conductivity only in the case of thin-film samples.
But many questions in the field of heavy fermions, in particular concerning quantum criticality, 
can only be addressed on bulk samples: the most interesting quantum-critical materials such as \YRS{} or CeCu$_{6-x}$Au$_x$ are not available so far as thin films. Furthermore there might be fundamental difficulties: for such samples, the mean free path of metallic electrons cannot exceed the film thickness. This sets the residual relaxation rate, which might then be too high to study intrinsic scattering processes like the Fermi-liquid behavior discussed below.
Therefore, we work on a completely different experimental approach that allows measurements on high-quality single crystals which are usually the core of experimental study of heavy-fermion quantum criticality.

This technique is based on planar superconducting resonators. Here we employ a stripline geometry (also called triplate), which is schematically illustrated in Fig.\ \ref{Fig_CoaxToPlanarScheme}: a flat strip acts as the center conductor of a microwave transmission line and is surrounded by conductive ground planes below and above. These three conductive layers are separated from each other by dielectrics, in our case by sapphire substrates.
While the center conductor has to be a conductive film that can be structured, the two ground planes only need to have one flat surface. Here, one of the ground planes is the single crystal sample which we want to study. Now, the damping of the microwave signal traveling along the line depends on the electrodynamic properties of the sample. However, a much stronger effect is given by the center conductor, where the microwave current density is much higher compared to the ground planes. To achieve that the sample as ground plane dominates the losses of the overall assembly, the center conductor is fabricated of a material with much lower losses than the sample under study, i.e.\ it is fabricated from a superconductor. At present, we usually work with Pb as superconducting material, but many other materials are possible as well. In principle one could use such a stripline arrangement for broadband measurements of the damping, i.e.\ from the measured transmission one could directly calculate the microwave properties of the sample. However, preliminary tests showed that the calibration of such an experiment is rather difficult. Therefore, we employ a different technique:\cite{DiIorio1988,Scheffler2012} by fabricating two gaps into the center conductor which reflect most of the microwave signal, we create a one-dimensional Fabry-Perot resonator. The resonator can be designed in such a way that its quality factor is governed by the sample of interest. Similar to the cavity resonators discussed above, one now measures the transmission through the resonator as a function of frequency, and from the width of the transmission maxima (like those shown in Fig.\ \ref{Fig_ESRstripline}(a)) one can directly calculate the quality factor. This does not depend on absolute values of the transmission, and when the resonances are very sharp in frequency space, these measurements are not susceptible to standing waves or frequency-dependent damping of the transmission line. 
Compared to cavity resonators, this approach has two main advantages: firstly, the sample geometry is very simple, which makes quantitative analysis comparably easy. Secondly, because we use a one-dimensional resonator, we can easily excite higher modes which lead to a set of resonances that are equally spaced in frequency. Compared to the broadband measurement, the number of discrete frequency points is small, but for many questions this might be sufficient as long as one covers a large frequency range: here we are most interested in features that are rather broad, e.g.\ the Drude response. In case that the resonance frequencies are spaced too far apart for a particular question, then one can easily implement further resonators with different resonator frequencies.

Besides the stripline approach, we also employ another planar resonator geometry, namely coplanar resonators. As schematically shown in Fig.\ \ref{Fig_CoaxToPlanarScheme}, a coplanar transmission line can conceptually be considered as a cut through a coaxial cable. In this geometry, inner and outer conductors are thin conductive layers, in our case superconducting Nb. The single-crystal heavy fermion sample is positioned at some distance above the resonator plane. A fraction of the microwave field penetrates the sample, and loss mechanisms in the sample affect the resonator $Q$. From the measured $Q$, one can then deduce microwave properties of the sample. The disadvantage of this approach is the difficult quantitative analysis, if absolute values of e.g.\ the microwave conductivity of the sample are wanted. (Coplanar line plus metallic sample form a non-trivial transmission line geometry.) But for ESR studies it is sufficient to measure relative changes as a function of magnetic field, and then the coplanar geometry has certain advantages compared to the stripline approach: firstly, one can confine the length of the resonator, which is required for a certain fundamental frequency, to a smaller surface area, making the device more compact. Secondly, the sample does not necessarily cover the complete resonator, i.e.\ smaller samples are sufficient. Finally, by adjusting the distance between resonator and sample, one can tune the total resonator $Q$ to a range that is convenient for straightforward measurements.

\section{Low-energy optics of heavy fermions}

If heavy fermions are excited with infrared light, one typically observes an absorption maximum that is often assigned to the \lq hybridization gap\rq{} and as such indicates an electronic excitation from the heavy-fermion state into an uncorrelated state. This has been studied in great detail for many heavy-fermion materials.\cite{Dordevic2001,Okamura2007} If instead one wants to optically address the heavy-fermion ground state, one has to work with much lower photon energies, i.e.\ THz and in particular microwave radiation. Here heavy fermions can act as well-defined model systems for certain aspects of the electronic properties of metals. Indeed, one of the cleanest observation of a simple Drude response was obtained on a heavy-fermion system with a relaxation rate of a few GHz, well below any other excitations.\cite{Dressel2006,Scheffler2005c,Scheffler2005b}

\subsection{Drude response}

In a first step, the optical properties of heavy fermions are usually discussed in the framework of the Drude description of metals.\cite{Dressel2002a} Its core concept is that the charge dynamics are governed by a single frequency scale, the transport relaxation rate $\Gamma$. This governs the frequency dependence of the optical conductivity: 
\begin{equation}
\sigma(\omega) = \sigma_1(\omega) + i \sigma_2(\omega) = \frac{\sigma_0}{(1-i\omega/\Gamma)} \, ,
\end{equation}
where $\sigma_0$ is the dc conductivity. This characteristic frequency dependence with a roll-off in $\sigma_1(\omega)$ and a maximum in $\sigma_2(\omega)$ at the relaxation rate is called Drude response. The relaxation rate contains information about the scattering processes of the charges, and therefore strongly depends on temperature, material, and sample quality. For conventional metals, $\Gamma$ is typically found at infrared frequencies. The large effective mass of heavy fermions leads to a suppression of $\Gamma$: the mass enhancement is equivalent to a reduction in the Fermi velocity, which in the low-temperature limit (where scattering is dominated by defects) leads to an equivalent decrease of the relaxation rate: $m^*/m = \Gamma / \Gamma^*$, where $m$ and $\Gamma$ are effective mass and relaxation rate of the uncorrelated system and $m^*$ and $\Gamma^*$ those of the correlated, heavy-electron system.\cite{Millis1987}
With a mass enhancement $m^*/m$ of up to 1000, the relaxation rate of heavy-fermion metals is shifted to frequency ranges much lower than the infrared, i.e.\ to the THz and microwave frequency range.

\begin{figure}[t]%
\includegraphics*[width=\linewidth]{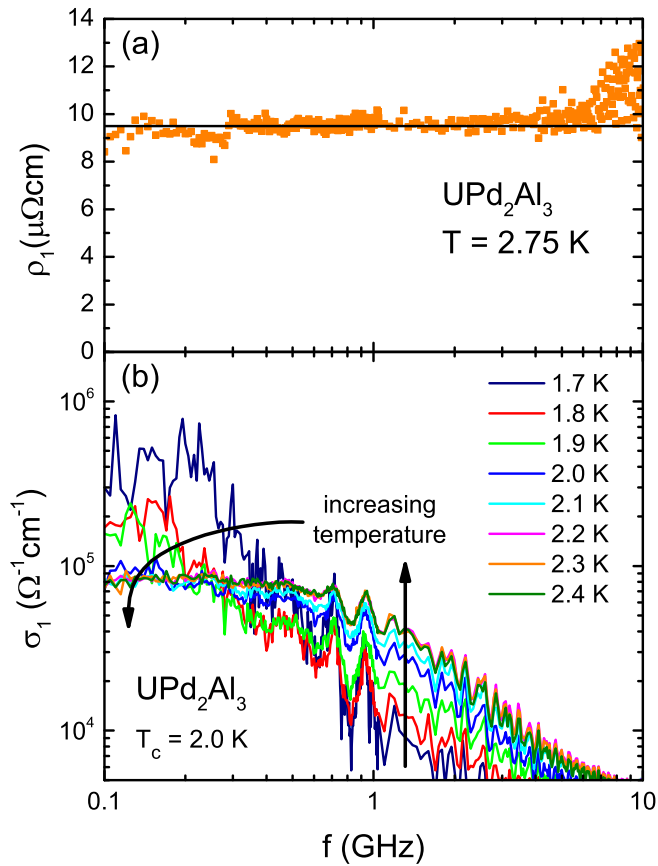}
\caption{%
 Frequency-dependent microwave properties of UPd$_2$Al$_3$ determined with Corbino spectroscopy. (a) Real part of the resistivity, which is proportional to the relaxation rate, in the heavy-fermion state above \Tc. The solid line is the $\omega^2$ expectation if the $T^2$ dependence of the dc resistivity were converted according to the Fermi-liquid prediction for the frequency dependence. (b) Real part of the conductivity above and below \Tc. Above \Tc, a pronounced Drude roll-off is observed, below \Tc{} a conductivity increase at low frequencies due to quasiparticles and suppressed conductivity at higher frequencies due to the energy gap in the density of states.}
\label{Fig_UPA}
\end{figure}

As an example, the conductivity spectra of \UPA{} for temperatures above the superconducting transition temperature \Tc, as shown in Fig.\ \ref{Fig_UPA}(b), display a pronounced roll-off above 1~GHz, which is the characteristic Drude response. So far, only two heavy-fermion materials, \UPA{} and \UNA, were studied with broadband microwave spectroscopy, and both showed Drude relaxation at GHz frequencies,\cite{Scheffler2006} whereas from infrared studies on other heavy-fermion compounds typically considerably higher relaxation rates are deduced. Future microwave studies should show whether the U-based compounds \UPA{} and \UNA{} are exceptional in their low relaxation rate or whether this is typical for other heavy-fermion materials as well.

\subsection{Fermi-liquid behavior}

The Drude response discussed so far considers the relaxation rate as a constant. Within the so-called extended Drude formalism, this assumption is lifted and the relaxation rate $\Gamma(\omega)$ is treated as a function of frequency.\cite{Dressel2002a} (To keep Kramers-Kronig consistency, also the effective mass $m$ becomes frequency dependent.)
Then, $\Gamma(\omega)$ can be calculated from experimental conductivity spectra $\sigma (\omega)$ as follows:
\begin{equation}\label{Eq_FreqDepGamma}
  \Gamma(\omega) = \frac{\omega_p^2}{4 \pi} {\rm Re}\left\{\frac{1}{\sigma(\omega)}\right\}
                 = \frac{\omega_p^2}{4 \pi} \frac{\sigma_1(\omega)}{|\sigma(\omega)|^2}
                 = \frac{\omega_p^2}{4 \pi} \rho_1(\omega) \, ,
\end{equation}
where $\rho_1$ is the real part of the complex resistivity $\rho = 1/\sigma$. $\omega_p = \sqrt{4 \pi n e^2 / m}$ is the unrenormalized plasma frequency of the material with charge carrier density $n$.\cite{Dressel2002a} Quantitative determination of $\omega_p$ of a heavy-fermion metal is a delicate task. Therefore we use the relation $\Gamma(\omega) \propto \rho_1(\omega)$ of Eq.\ \ref{Eq_FreqDepGamma} and discuss $\Gamma$ as the fundamental, microscopic quantity and $\rho_1 = {\rm Re}(1/\sigma)$ as the corresponding quantity which is directly determined from experiment.

The extended Drude formalism allows a generalization of the metallic response, going from the simple Drude case, which considers the charge carriers as non-interacting, to the more complex case of electronic interactions. The canonical framework to describe electronic interactions in metals is Landau's theory of Fermi liquids.
Fermi liquid (FL) theory predicts that the transport relaxation rate depends both on temperature $T$ and frequency $\omega$. In the presence of an additional residual relaxation rate $\Gamma_0$ (due to defects), one expects:
\begin{equation}
\label{Eq_FL}
\Gamma(T,\omega) = \Gamma_0 + a (k_B T)^2 + b (\hbar \omega)^2
\end{equation}
where $a$ and $b$ are material-specific constants, and for a conventional, realistic case of a simple Fermi liquid: $a/b = 4\pi^2$.\cite{Dressel2002a,Gurzhi1959,Degiorgi1999,Rosch2005}
In fact, the $T^2$-behavior, which can be accessed by temperature-dependent measurements of the dc resistivity $\rho_{dc} = \rho_0 + A (k_B T)^2$, is typically taken as the hallmark whether a metal is in the FL regime or not. (According to Eq.\ \ref{Eq_FreqDepGamma}, the prefactors $A$ and $B$ of FL behavior in $\rho$ and those, $a$ and $b$, for $\Gamma$ are simply related: $A/a = B/b = 4\pi/\omega_p^2$.) While for a conventional metal it is difficult to observe this $T^2$ behavior (it is usually covered by scattering due to phonons and defects), this is comparably simple for heavy fermions: the FL prefactors $a$ and $b$ depend quadratically on effective mass and thus the FL contributions to the transport relaxation rate of heavy fermions are very strong. This is consistent with the numerous cases where $T^2$ behavior was observed in the dc resistivity of heavy fermions.

The situation is quite different for the $\omega^2$ dependence. Although there are experimental observations of the optical scattering rate $\Gamma(\omega)$ with frequency dependence $\propto \omega^2$,\cite{Dordevic2006} none of them should be considered as conclusive demonstration of FL behavior: most of these experiments were performed at infrared frequencies, i.e.\ energy scales much higher than usually considered as FL regimes. Furthermore, there is no unequivocal case where $\omega^2$ and $T^2$ behavior was observed in consistent temperature and frequency ranges, let alone the predicted prefactor ratio $a/b$.\cite{Dressel2011}

In recent years, this unsatisfying situation regarding the $\omega^2$ dependence in FL optics has become more and more a concern for the scientific community studying optics of metals. This lead to new efforts addressing the $\omega^2$ behavior of Fermi liquids. While theoretical work focuses on how the plain prediction by Gurzhi, Eq.\ \ref{Eq_FL},\cite{Gurzhi1959} should be interpreted and applied to realistic cases of present interest,\cite{Chubukov2012,Maslov2012} new experimental work consists of a more detailed assessment of the found $\omega^2$ data and a scrutinizing discussion in the context of Fermi liquids. But at least from the experimental side, no consensus was obtained yet.\cite{Dressel2011,Nagel2011}

To finally settle the question how the $\omega^2$ FL behavior manifests experimentally, heavy-fermion materials are an appropriate choice: many heavy-fermion materials are well-established Fermi liquids, and with their enhanced $a$ and $b$ prefactors, the FL contributions to the optical response should be very strong. Furthermore, for certain heavy fermions close to a quantum-critical point, the $A$ prefactor was shown to increase and possibly even diverge when one approaches the quantum-critical point by adjustment of an external control such as magnetic field or pressure.\cite{Gegenwart2002,Paglione2003} In these cases, heavy fermions can be considered as continuously tunable metals, something rarely found in nature. For the $\omega^2$ question this offers the opportunity to choose a material that is known to be a Fermi liquid in certain regimes of its phase diagram. If one finds $\omega^2$ behavior in these phases and if the $A$ and $B$ prefactors of $\rho_1(T,\omega)= \rho_0 + A (k_B T)^2 + B (\hbar \omega)^2$ scale with each other upon tuning the external parameter, then one has found very solid realization of optical FL behavior, and such an experiment will finally solve the question of whether the ratio $A/B$ amounts to 4$\pi^2$.
When considering such an experiment, one has to carefully choose the appropriate frequency range. Clearly, a larger frequency will lead to a much stronger $\omega^2$ contribution, but the frequency has to be low enough to ensure that one addresses the heavy-fermion ground state. For several U-based heavy fermions, THz measurements suggest that at least for these compounds one observes the heavy-mass quasiparticles only for frequencies well below 100~GHz.\cite{Ostertag2011,Donovan1997,Dressel2002b} Therefore, microwave frequencies are the obvious spectral range to search for FL optics in heavy fermions.

In the extended Drude analysis, $\rho_1$ is proportional to the relaxation rate $\Gamma$. Fig.\ \ref{Fig_UPA}(a) shows that $\rho_1$ of the heavy-fermion state in \UPA{} is basically constant as a function of frequency. (The increase at the highest frequencies is not significant within the experimental error.)\cite{Scheffler2010} \UPA{} has a quadratic temperature dependence of the dc resistivity $\rho_{dc} = \rho_0 + A T^2$ in a limited temperature range above \Tc.\cite{Dalichaouch1992} It is not clear whether for this particular material this indicates FL behavior, but if one calculates the corresponding frequency dependence (following Eq.\ \ref{Eq_FL}), then one obtains the line in Fig.\ \ref{Fig_UPA}(a), which is basically flat in this frequency range. Even at 20~GHz, the increase in $\rho_1$ compared to the dc value is only $6*10^{-5}$ and thus far too small to be observed. For comparison, the corresponding FL response of \YRS{} should exceed 20\% if a microwave experiment could be performed at 20~GHz, at 100~mK, and in a tunable magnetic field.

\subsection{Quantum criticality}

The non-Fermi-liquid (NFL) properties of heavy fermions close to a quantum-critical point \cite{vLoehneysen2007,Gegenwart2008} are another obvious question for optical studies. While there are several infrared studies of heavy fermions which are discussed in terms of NFL behavior and quantum criticality, one again should ask the question of the appropriate frequency and temperature ranges. To be able to directly compare frequency-dependent conductivity data to well established NFL properties of heavy fermions, which are typically confined to temperatures of order 1~K or below, microwave experiments at mK temperatures are desired.
One of the main questions to be addressed here is $\omega/T$ scaling. In a seminal work more than a decade ago, Schroeder \textit{et al.} studied CeCu$_{6-x}$Au$_x$ with neutron spectroscopy and they found in the dynamical susceptibility non-trivial $\omega/T$ scaling as a strong indication of unconventional quantum criticality.\cite{Schroeder2000} Since then there have been more indirect studies of $\omega/T$ scaling for quantum-critical heavy fermions,\cite{Aynajian2012,Friedemann2010} whereas so far there is no corresponding optical study, although this would offer another direct access to $\omega/T$ as an independent energy scale can be compared to temperature.

\begin{figure}[t]%
\includegraphics*[width=\linewidth]{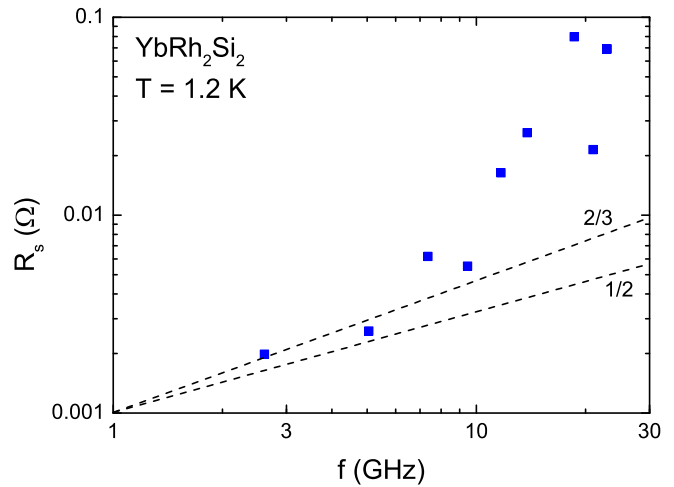}
\caption{%
 Surface resistance of YbRh$_2$Si$_2$ at 1.2~K. Data were obtained from measurements of a single-crystalline YbRh$_2$Si$_2$ sample in a superconducting Pb stripline resonator. Dashed lines correspond to power laws with exponent 1/2 (normal skin effect) and 2/3 (anomalous skin effect).}
\label{Fig_YRSstripline}
\end{figure}

In Fig.\ \ref{Fig_YRSstripline} we show microwave data of \YRS{} at a temperature of 1.2~K and without applied magnetic field, i.e.\ in the NFL regime of the material.\cite{Trovarelli2000} From the resonator transmission spectra, we obtain the resonance frequency and quality factor, which allows the determination of the surface resistance $R_s$ of the sample material. Clearly, $R_s$ increases with frequency as expected for a metal. For comparison, the dashed lines Fig.\ \ref{Fig_YRSstripline} indicate power-law frequency dependences with exponents 1/2 and 2/3, corresponding to the normal and anomalous skin effects, respectively,\cite{Dressel2002a} both of which might be realized in this metal at cryogenic temperatures. But the experimental $R_s$ data of \YRS{} increase much more strongly with frequency, which indicates a non-Drude like conductivity spectrum, which is consistent with the pronounced NFL behavior of this quantum-critical material in dc resistivity for temperatures as high as 10~K.
This experiment can easily be transferred to a dilution refrigerator and to magnetic fields of the order of 100~mT, and microwave studies concerning the electrodynamic properties of \YRS{} close to the quantum-critical point become feasible.

\subsection{Superconducting heavy fermi\-ons}

So far, we only discussed the microwave response in the metallic state. But both heavy-fermion materials studied with the Corbino technique, UPd$_2$Al$_3$ and UNi$_2$Al$_3$, undergo superconducting transitions, at \Tc{} = 2.0~K and \Tc{} = 1.0~K, respectively.\cite{Geibel1991a,Geibel1991b} Corbino spectroscopy has already been used to study the microwave response of high-\Tc{} and low-\Tc{} superconductors,\cite{Booth1996a,Ohashi2006,Steinberg2008,Pompeo2010,Liu2011} and it is an obvious next step to address heavy-fermion superconductors. In the superconducting state, the impedance of the sample is much smaller compared to the metallic state, and thus the resolution of a Corbino spectrometer might not be sufficient to resolve the superconducting state. In Fig.\ \ref{Fig_UPA}(b) we show Corbino measurements on the same sample studied in Refs.\ \cite{Scheffler2005c,Scheffler2010} for the normal state, but now below \Tc. The data were obtained in a $^4$He cryostat and it was calibrated using a short-only procedure with a bulk aluminium sample as short.\cite{Scheffler2005a}
In Fig.\ \ref{Fig_UPA}(b) one clearly sees the electrodynamic signature of the superconducting transition:\cite{Dressel2002a} at higher frequencies, above 0.5~GHz, the real part $\sigma_1$ of the conductivity is suppressed, indicating the development of the superconducting energy gap. At frequencies below 0.3~GHz, on the other hand, $\sigma_1$ increases drastically, which at these temperatures (1.7~K $\leq$ T $\leq$ T$_c$ = 2.0~K) is due to the presence of thermally excited quasiparticles.\cite{Steinberg2008,Turner2003} The increase of data scattering in the conductivity spectra for the lower temperatures, which in fact is not random noise but systematic errors not compensated for by the calibration, is directly related to the impedance decrease.\cite{Scheffler2005a} For the sample presented here, going to temperatures much below \Tc{} leads to measured spectra where the error in the signal is larger than the signal itself. This problem can be overcome by using the strip geometry.\cite{Scheffler2007} Combining this with temperatures below 1~K, which become accessible with recently developed Corbino spectroscopy at $^3$He temperatures,\cite{Liu2011,Steinberg2012} even the superconducting state of \UNA{} becomes accessible. Another candidate material for broadband studies of superconducting heavy fermions are thin films of CeCoIn$_5$.\cite{Mizukami2011}

\begin{figure}[t]%
\includegraphics*[width=\linewidth]{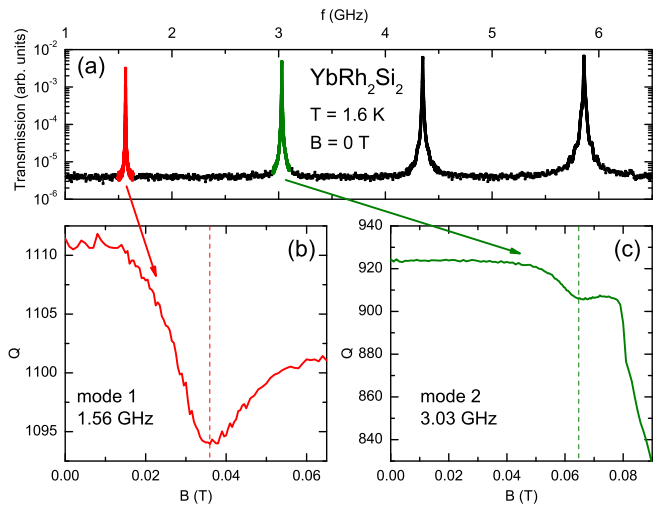}
\caption{%
 (a) Transmission spectrum of a superconducting Pb stripline resonator with \YRS{} sample. The first four resonant modes are shown.
 (b), (c) Quality factor of the first two resonant modes as a function of external magnetic field (applied parallel to the superconducting resonator and perpendicular to the crystallographic $c$-axis of the \YRS{} sample). ESR is manifested by the minima in $Q$, indicated by dashed lines.}
\label{Fig_ESRstripline}
\end{figure}

\section{ESR on heavy fermions}

Up to this point we were concerned with the charge response of heavy fermions at microwave frequencies, i.e.\ we probe the charge dynamics by applying a high-frequency electric field. In the following, we describe a rather different approach, that is probing the dynamics of magnetic moments by applying a high-frequency magnetic field. For almost all conditions, this magnetic response is orders of magnitude weaker than the charge response and thus not accessible by experiment. The exception being the occurrence of magnetic resonance phenomena: if a static magnetic field is applied to a material with magnetic moments (local moments or conduction electron moments), then the precession motion of the moment around the field direction can be probed by an oscillating (microwave) magnetic field. If static magnetic field $H$ and microwave frequency $f = \omega/(2\pi)$ fulfill the resonance condition
\begin{equation}
 h f = g \mu_B H
\end{equation}\label{Eq_ESR}
(with $h$ Planck's constant and $\mu_B$ the Bohr magneton), then resonance absorption occurs that can be detected. For the case of paramagnetic materials, this magnetic resonance is called electron paramagnetic resonance (EPR) or electron spin resonance (ESR), respectively. ESR is a common technique e.g.\ in chemistry, but it can also help to understand the spin properties of complex electron systems in solids.
But in heavy-fermion systems local moments, which are fundamental to the development of the heavy-fermion state, were expected to be fully screened by the conduction electrons. Thus, heavy-fermion compounds were considered to not display any measurable ESR signals unless ESR active probe spins (such as those of Gd$^{3+}$) are doped and give indirect information on the heavy-fermion spin dynamics.\cite{Krug1998}
This conception changed drastically when a pronounced ESR signal was observed in \YRS.\cite{Sichelschmidt2003} Since then, ESR of \YRS{} has been studied in great detail.\cite{Sichelschmidt2003,Duque2009,Schaufuss2009,Sichelschmidt2010}  One particularly interesting result is that ESR $g$ factor and linewidth may reflect the evolution of the heavy FL state throughout the $B$-$T$-phase diagram of this quantum-critical material.\cite{Schaufuss2009,Sichelschmidt2010} 
However, although many different ESR setups were employed, the most interesting regions of the phase diagram, including the antiferromagnetic phase (at temperatures below 70~mK and in-plane magnetic fields below 60~mT) and the adjacent quantum-critical as well as Fermi-liquid regimes, could not be addressed so far. This has two reasons: firstly, conventional ESR spectrometers operate at a fixed frequency, with two particularly common frequencies being 9.4~GHz (X band) and 34~GHz (Q band). In the case of \YRS, this corresponds to ESR resonance fields of approximately 200~mT and 700~mT.	These fields are considerably higher than the quantum-critical field. Furthermore, a general trend of ESR spectroscopy is to develop techniques that work at ever higher frequencies, because a larger ESR resonance field allows better resolution.\cite{Takahashi2012} In contrast, here we are interested in ESR resonance fields in the range 10~mT to 300~mT, and it is desirable to be able to adjust the ESR frequency to the magnetic field strength of interest. Furthermore, temperatures below 100~mK are required, i.e.\ one has to perform ESR in a dilution refrigerator. While some ESR experiments were performed at mK temperatures since the 1970s, none of them is directly applicable to our special case of interest with ESR frequencies of a few GHz and metallic samples. So far, the lowest temperatures of ESR studies on \YRS{} with conventional ESR spectrometers are $^3$He temperatures down to 500~mK.\cite{Sichelschmidt2010}

\begin{figure}[t]%
\includegraphics*[width=\linewidth]{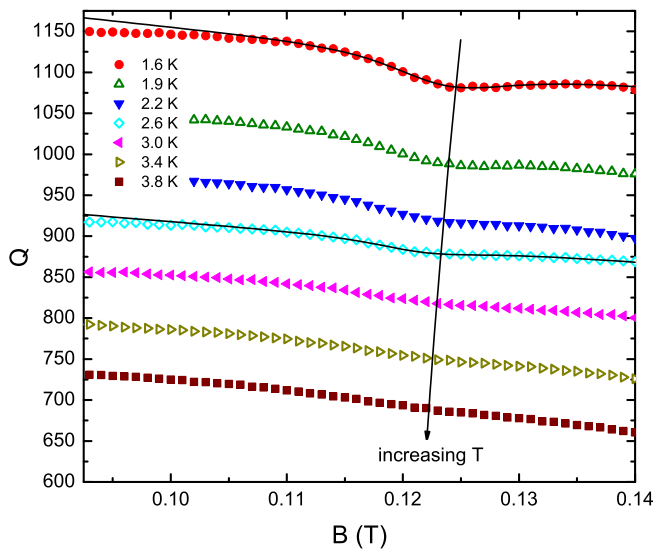}
\caption{%
 Quality factor of superconducting coplanar resonator (5.97~GHz) with YbRh$_2$Si$_2$ sample. ESR is manifested as minimum in the field dependence, which weakens and shifts toward lower magnetic field upon warming, as indicated by the arrow. For two ESR spectra, the Lorentzian fit curves are also shown.}
\label{Fig_ESRcoplanar}
\end{figure}

\subsection{ESR of YRh$_2$Si$_2$ using planar resonators}

In an ongoing study, we try to overcome these experimental limitations by using planar superconducting resonators. In one approach, we employ stripline resonators similar to those discussed in section \ref{SectionPlanarResonators}. In these stripline resonators, in addition to the electrical field component of the microwave signal that we discussed above, there is also a microwave magnetic field present. By adjusting the geometry of the inner conductor with respect to an externally applied static magnetic field (static and microwave magnetic fields perpendicular to each other), we can achieve a field geometry that supports ESR for a substantial part of the sample surface.
In Fig. \ref{Fig_ESRstripline}(b) and (c) we show exemplary ESR spectra for \YRS: the quality factor $Q$ of the first two modes of a superconducting Pb stripline resonator is plotted as a function of static magnetic field. ESR is manifested as the minima in $Q$ around 35~mT for the first mode and 65~mT for the second mode. For the second mode, the ESR line is a signal on top of a field-dependent background of the superconducting resonator; here the steep drop in $Q$ at the critical field of Pb around 80~mT is particularly pronounced. For a detailed analysis of the ESR absorption line, this background has to be modeled carefully.

Coplanar resonators can also be employed for ESR studies.
In Fig.\ \ref{Fig_ESRcoplanar} we show exemplary ESR spectra of a \YRS{} sample mounted on top of a coplanar Nb resonator. At low temperatures, such as 1.6~K, the ESR of \YRS{} is clearly visible as a dip in the quality factor. With increasing temperature, the ESR becomes weaker. As with the Pb stripline, the superconducting resonator itself has a field dependence which in this case of Nb is much more gradual. Understanding this field dependence of superconducting coplanar resonators in a magnetic field and approaches to inhibit the suppression of $Q$ are topics of active research, including in the field of solid-state quantum information processing.\cite{Song2009,Bothner2012,Ranjan2012}

\begin{figure}[t]%
\includegraphics*[width=\linewidth]{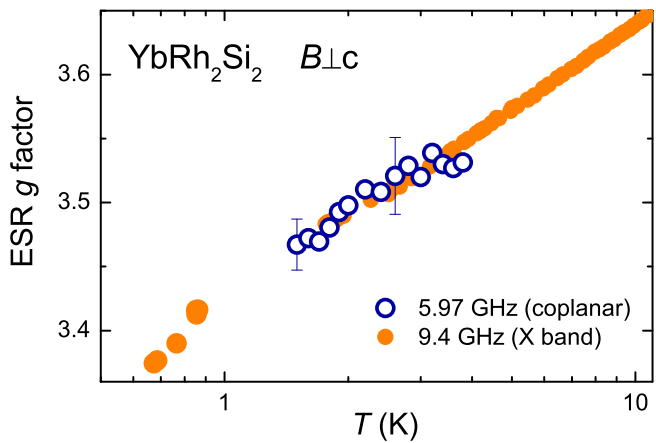}
\caption{%
 ESR $g$ factor of YbRh$_2$Si$_2$ determined from the ESR spectra in Fig.\ \ref{Fig_ESRcoplanar}. For comparison, the $g$ factor obtained from conventional X band ESR measurements is also displayed. (Representative error bars are shown for two data points.)}
\label{Fig_gFactor}
\end{figure}

The ESR $g$ factor of \YRS, which we obtain from fitting each ESR spectrum with a Lorentzian shape (Fig.\ \ref{Fig_ESRcoplanar}), is shown in Fig.\ \ref{Fig_gFactor}. For comparison, we also plot the result of previous ESR measurements on \YRS{} using conventional X band spectrometers.\cite{Sichelschmidt2010} Although the frequencies and the corresponding ESR resonance fields differ slightly, the obtained $g$-factors agree well. The larger error bars of our new measurements are due to the difficulties to properly describe the field-dependent background of the superconducting resonator. A better understanding of this background should lead to a substantial reduction of these error bars.
Already at this experimental status, ESR measurements at mK temperatures are very promising: both, the stripline and the coplanar resonators can easily be implemented in a commercial dilution refrigerator. Furthermore, as the ESR absorption becomes stronger upon cooling, the role of the field-dependent background of the superconducting resonators becomes less important.

\subsection{ESR on quantum-critical heavy fermions}

With this new experimental approach, we expect to explore a major fraction of the most interesting regions of the $B$-$T$-phase diagram of \YRS{} in the vicinity of its quantum-critical point. Here we would like to address the following questions: 

What is the nature of ESR in Kondo lattices? We still do not have a good understanding of why ESR is present in certain heavy-fermion materials. Of particular interest here is a study over a wide temperature range, preferably exclusively in the quantum-critical regime, without crossing any boundaries of the phase diagram. Already the first ESR studies on \YRS{} indicated that the ESR $g$-factor might be tightly connected to the susceptibility with its characteristic logarithmic temperature dependence, and this might be corroborated by measurements over several orders of magnitude in temperature. A very fundamental question in this regard is whether the ESR in \YRS{} should be interpreted in the framework of local moments or of Landau quasiparticles of Fermi liquids, which could be extended into the NFL regime.\cite{Kochelaev2009,Woelfle2009}

How does ESR reflect the phase diagram of \YRS? From experimental work, there are indications that the temperature dependence of ESR $g$-factor and linewidth changes at the so-called $T^*$ line (a transition between small and large Fermi surface).\cite{Schaufuss2009,Sichelschmidt2010} On the other hand, from the theoretical side it is predicted that for ESR caused by Landau-type quasiparticles the transition from the NFL to the FL regime should affect the ESR properties, in particular the temperature dependence of $g$-factor and the linewidth.\cite{Woelfle2009}

How does ESR evolve in the antiferromagnetic state of \YRS? So far, the antiferromagnetic state of \YRS{} at temperatures below 70~mK and in-plane fields of 60~mT is by far not studied in as much detail as desired. At this stage, the antiferromagnetic order is thought to be of local nature,\cite{Paschen2004} in contrast to the possibility of a spin density wave. But there is basically no spectroscopic evidence concerning the antiferromagnetic state. Most desired would be neutron scattering, which has been of great help for the understanding of other quantum-critical heavy-fermion systems,\cite{Schroeder2000} but is notoriously difficult for \YRS.\cite{Stock2012}
We expect that the ESR signal of \YRS{} changes considerably upon entering the antiferromagnetic phase. For the case of a local-moment antiferromagnet, the ESR should modify to an antiferromagnetic resonance (AFMR), where the resonance field is shifted compared to the paramagnetic case due to the internal fields caused by the magnetic sublattices of the antiferromagnet. How strong this effect will be remains to be seen, in particular since the local moment of the antiferromagnetic state is very small.

So far, we discussed these ESR-related questions concerning quantum-critical heavy fermions only with respect to \YRS. But there are more materials that are of interest in this regard, e.g.\ doped systems derived from \YRS. Several of them have already been studied with ESR at temperatures above 2~K, but many of them are of large interest at mK temperatures as well, because the electronic ground state can change drastically when \YRS{} is doped. Of particular interest here is the putative quantum spin liquid state.\cite{Friedemann2009}
A rather different material is $\beta$-YbAlB$_4$, a heavy-fermion material that exhibits superconductivity below 80~mK and pronounced signatures of quantum criticality in the $B$-$T$-phase diagram.\cite{Nakatsuji2008} In particular, the magnetic field of the quantum-critical point is very close to zero field.\cite{Matsumoto2011}
ESR was found in $\beta$-YbAlB$_4$ at temperatures up to room temperature, with a pronounced change of $g$-factor as well as the appearance of hyperfine splitting upon cooling.\cite{Holanda2011} These measurements were performed at 9.5~GHz and temperatures above 4.2~K. Clearly, it is of interest to go to much lower temperatures and fields, to explore with ESR the phase diagram of $\beta$-YbAlB$_4$ with FL, NFL, and superconducting phases, and to study the role that quantum criticality plays for the ESR in this compound, which can lead to complimentary results compared to those obtained on \YRS.

\section{Conclusions}

Microwave spectroscopy probes electrical charges via interaction with the electrical field component and magnetic moments via interaction with the magnetic field component of the high-frequency radiation. The energy scales of microwave radiation correspond to temperatures of 1~K and below, and as such are well suited to address the characteristic excitations and dynamics of heavy-fermion systems. We have shown that broadband microwave spectroscopy allows the study of the slow Drude response of heavy fermions as well as the charge dynamics in the superconducting state. For the ongoing discussions about quantum criticality in the vicinity of a quantum phase transition, broadband microwave spectroscopy is not applicable at the moment because of the lack of appropriate thin film samples of quantum-critical heavy fermions. Therefore, we suggest planar superconducting resonators as an alternative approach that is compatible with mK temperatures and capable of resolving the electrodynamic response of heavy-fermion metals close to a quantum phase transition. A very similar experiment can be used for ESR studies, as demonstrated for \YRS. With these techniques available, previously impossible studies of heavy fermions at very low temperature have become feasible, and several questions connected to quantum criticality as well as heavy-fermion superconductivity can be addressed.

\section*{Acknowledgements}
We thank G.\ Untereiner, D.\ Bothner, R.\ Kleiner, and D.\ K\"olle for support with resonator fabrication.
We acknowledge helpful discussions with D.\ Broun, A.\,V.\ Chubukov, A.\ Georges, S.\ Kirchner, B.\,I.\ Kochelaev, D.\,L.\ Maslov, S. Paschen, and A.\ Schofield.
Financial support by the DFG, including SFB/TRR21, and the Volkswagen Stiftung (I/84689) is thankfully acknowledged.

\end{document}